\documentclass[aps,prl,nofootinbib,twocolumn,superscriptaddress,floatfix,notitlepage]{revtex4-1} 

\usepackage{graphicx}
\usepackage{amsmath,amsfonts,amssymb}
\usepackage{color}
\usepackage[breaklinks,colorlinks,urlcolor=blue,citecolor=magenta,linkcolor=magenta,hypertexnames=false]{hyperref}
\usepackage{verbatim}

\newcommand{\be}{\begin{equation}} 
\newcommand{\ee}{\end{equation}}
\newcommand{\bea}{\begin{equation}\begin{aligned}} 
\newcommand{\eea}{\end{aligned}\end{equation}}

\newcommand{\td}{{\rm d}}

\newcommand{\papertitle}{Black holes and gravitational waves from slow first-order phase transitions}

\begin{document}

\title{\papertitle}

\author{Marek Lewicki}
\email{marek.lewicki@fuw.edu.pl}
\affiliation{Faculty of Physics, University of Warsaw, ul. Pasteura 5, 02-093 Warsaw, Poland}

\author{Piotr Toczek}
\email{piotr.toczek@fuw.edu.pl}
\affiliation{Faculty of Physics, University of Warsaw, ul. Pasteura 5, 02-093 Warsaw, Poland}

\author{Ville Vaskonen}
\email{ville.vaskonen@pd.infn.it}
\affiliation{Dipartimento di Fisica e Astronomia, Universit\`a degli Studi di Padova, Via Marzolo 8, 35131 Padova, Italy}
\affiliation{Istituto Nazionale di Fisica Nucleare, Sezione di Padova, Via Marzolo 8, 35131 Padova, Italy}
\affiliation{Keemilise ja bioloogilise f\"u\"usika instituut, R\"avala pst. 10, 10143 Tallinn, Estonia}

\begin{abstract}
Slow first-order phase transitions generate large inhomogeneities that can lead to the formation of primordial black holes (PBHs). We show that the gravitational wave (GW) spectrum then consists of a primary component sourced by bubble collisions and a secondary one induced by large perturbations. The latter gives the dominant peak if $\beta/H_0 < 12$, impacting, in particular, the interpretation of the recent PTA data. The GW signal associated with a particular PBH population is stronger than in typical scenarios because of a negative non-Gaussianity of the perturbations and it has a distinguishable shape with two peaks.
\end{abstract}

\maketitle

\vspace{5pt}\noindent\textbf{Introduction --} In addition to the primordial inflation during which the fluctuations seen in the CMB were generated, the early Universe may have experienced another period of inflation. A strongly supercooled phase transition (PT) may have caused such period of thermal inflation if the vacuum energy density of the false vacuum dominated over the radiation energy density before the transition occurred. If the transition was also slow, this may have led to the formation of large inhomogeneities that sourced both primordial black holes (PBHs) and gravitational waves (GWs).

PBHs can form either during or after the transition. For the former case, the transition needs to be of first order, proceeding by bubble nucleation, and PBHs form from subhorizon patches that are surrounded by the bubble walls~\cite{Hawking:1982ga, Kodama:1982sf, Lewicki:2023ioy}. In the latter case, the transition can be either a continuous transition~\cite{Dimopoulos:2019wew} or a first-order PTs~\cite{Liu:2021svg, Kawana:2022olo, Gouttenoire:2023naa, Baldes:2023rqv} and PBHs form when the large fluctuations generated in the transition re-enter horizon. 

This letter focuses on the formation of large inhomogeneities during a slow and strongly supercooled first-order PT. During such a process each Hubble patch includes only a handful of large bubbles and the perturbations originate from the fluctuations in the times of their nucleation. We compute the distribution of the perturbations using a new semi-analytic approach and the PBH formation following the standard formalism~\cite{Carr:2020gox}.

First-order PTs generate also a GW background~\cite{Caprini:2015zlo,Caprini:2019egz}. For strongly supercooled cases, generic in quasi-conformal models~\cite{Jinno:2016knw, Iso:2017uuu, Marzola:2017jzl, Kierkla:2022odc, Kierkla:2023von, Prokopec:2018tnq, Marzo:2018nov, Baratella:2018pxi, VonHarling:2019rgb, Aoki:2019mlt, DelleRose:2019pgi, Wang:2020jrd, Ellis:2020nnr, Baldes:2020kam, Baldes:2021aph, Lewicki:2021xku, Gouttenoire:2023pxh}, it is sourced by the collisions of the bubble walls and relativistic fluid shells~\cite{Kosowsky:1992vn,Lewicki:2022pdb}. For slow transitions, there is also another source: the large perturbations. These GWs arise from the second-order terms in cosmological perturbation theory and their spectrum is widely studied in the literature~\cite{Domenech:2021ztg}. In this letter, we show, for the first time, that the induced GW background is stronger than that from the bubble collisions if the transition is sufficiently slow. In particular, this is the case for transitions that lead to PBH formation.

The induced GW background is a signature also in the typical PBH formation scenarios, such as the constant-roll inflation~\cite{Kannike:2017bxn, Germani:2017bcs, Motohashi:2017kbs, Karam:2022nym}. However, we find that the distribution of the perturbations generated in the first-order PT has a negative non-Gaussianity that suppresses the PBH formation. Consequently, the GW background associated with a certain abundance of PBHs, in this case, is stronger than e.g. in the case of constant-roll inflation where the non-Gaussianity is positive~\cite{Tomberg:2023kli}.

\vspace{5pt}\noindent\textbf{Formation of inhomogeneities --} We consider an exponential bubble nucleation rate
\be
	\Gamma(t) = H_0^4 e^{\beta t} \,,
\ee
where $H_0$ denotes the Hubble rate at $t=0$ and $\beta>0$ parametrizes the growth of the nucleation rate. The transition is slow if $\beta/H_0 = \mathcal{O}(1)$ and the time it takes for the averaged false vacuum fraction $\bar{F}(t)$~\cite{Guth:1982pn} to get from $\bar{F}(t)\approx 1$ to $\bar{F}(t)\approx 0$ is comparable to the Hubble time $1/H_0$.

During the period of thermal inflation the comoving Hubble horizon radius $(aH)^{-1}$ shrinks and reaches its smallest value roughly at the percolation time $t_p$, defined as $\bar{F}(t_p) = 0.3$. The corresponding largest comoving wavenumber that exits horizon can be approximated as $ k_{\rm max} = a(t_p) H(t_p)$, giving
\be \label{eq:kmax}
    k_{\rm max} \approx 1.6\times 10^{-7} {\rm Hz} \,\bigg[ \frac{g_*}{100} \bigg]^{\frac12} \bigg[ \frac{100}{g_{*s}} \bigg]^{\frac13} \frac{T_{\rm reh}}{\rm GeV} \,,
\ee
where $T_{\rm reh} \approx 8.5\times 10^8\,{\rm GeV} [g_*/100]^{-1/4} \sqrt{H_0/{\rm GeV}}$ is the temperature that the Universe reaches right after the thermal inflation and $g_*$ and $g_{*s}$ are the effective numbers of relativistic energy and entropy degrees of freedom at $T=T_{\rm reh}$.

The expanding bubbles convert the vacuum energy into kinetic and gradient energies of the bubble walls. The energy density in the walls scales in the same way as that of radiation~\cite{Lewicki:2023ioy} and eventually decays to radiation after the bubble collisions. In total, the energy density in the bubble walls and radiation evolves as
\be \label{eq:rhor}
    \dot\rho_r + 4 H \rho_r = -\dot\rho_v \,,
\ee
where dot denotes the time derivative and $\rho_v$ is the vacuum energy density. The background false vacuum energy density is given by $\bar\rho_v(t) = \bar{F}(t) \Delta V$, where $\Delta V$ denotes the vacuum energy density difference between the true and false vacua, and the Hubble rate by $H^2 = 8\pi G (\bar\rho_r + \bar\rho_v)/3$, where $\bar\rho_r$ is the solution of Eq.~\eqref{eq:rhor} with $\rho_v(t) = \bar\rho_v(t)$. The evolution of the scale factor is determined by $\dot a = aH$.

On superhorizon scales each comoving Hubble patch evolves mutually independently and the false vacuum fraction evolves differently in different patches. We compute the false vacuum fraction $F_k(t)$ in different patches using the semi-analytical method derived in the Supplemental Material~\cite{SM}. This accurately accounts for the fluctuations in the nucleation times and positions of the first $50$ bubbles and averages over the rest of the bubbles. For given $\beta/H_0$ and scale $k$, we generate $10^6$ realizations of the false vacuum fraction $F_k(t)$, solve the radiation energy density $\rho_{r,k}(t)$ from Eq.~\eqref{eq:rhor} with $\rho_v(t) = F_k(t) \Delta V$ and compute the density contrast as
\be
    \delta = \frac{\rho_k(t_k)}{\bar\rho(t_k)} - 1\,,
\ee
where $\rho_k(t) = \rho_{r,k}(t) + F_k(t) \Delta V$ and $\bar\rho(t) = \bar\rho_r(t) + \bar\rho_v(t)$. We evaluate the density contrast at the time $t = t_k \geq t_p$ when the scale $k$ re-enters horizon, $a(t_k) H(t_k) = k$.

\begin{figure}
\centering
\includegraphics[width=\columnwidth]{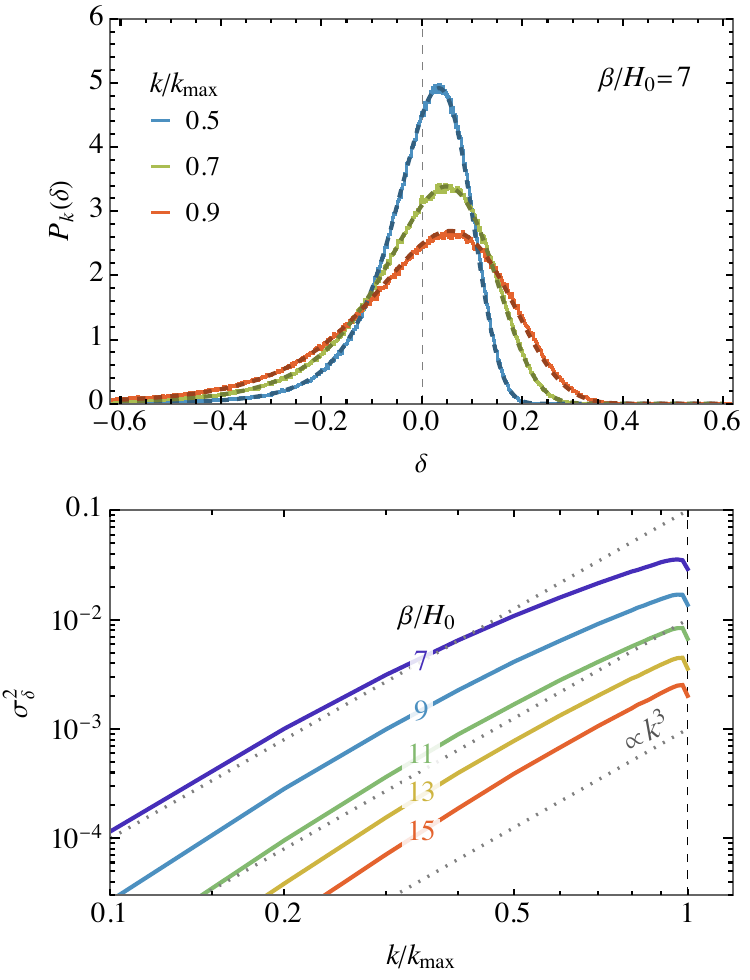}
\caption{\emph{Upper panel:} The probability distribution of the density contrast at different wavenumbers $k$. The histograms show the distributions of $10^6$ realizations and the dashed curves the fits using Eq.~\eqref{eq:Pkfit}. \emph{Lower panel:} The variance of the density contrast as a function of $k$ for different $\beta/H_0$ values.}
\label{fig:delta}
\end{figure}

We show the probability distribution of $\delta$ in the upper panel of Fig.~\ref{fig:delta}. The distributions are non-Gaussian with an exponential tail at negative $\delta$ and rapid fall at positive $\delta$. We find that the distribution can be fitted with
\be \label{eq:Pkfit}
    P_k(\delta) \propto \exp\left[ \frac{\epsilon}{2} (\delta-\mu) - \frac{2}{\epsilon^2 \sigma^2} \!\left( 1 - e^{\frac{\epsilon}{2} (\delta-\mu)} \right)^{\!2} \right] ,
\ee
where $\epsilon, \sigma > 0$. This distribution with $\epsilon < 0$ was found in models with constant roll inflation~\cite{Tomberg:2023kli}. In the lower panel of Fig.~\ref{fig:delta}, we show the spectrum of $\delta$ as a function of $k$. We cut the spectrum at $k=k_{\rm max}$ as the smaller length scales don't exit horizon during the thermal inflation. The production of large density fluctuations in a setup not including a period of thermal inflation was recently estimated in~\cite{Liu:2022lvz,Elor:2023xbz}.

\vspace{5pt}\noindent\textbf{Primordial black holes --} Large density perturbations lead to formation of PBHs as the radiation pressure is not enough to prevent the collapse of the overdense patch when it re-enters horizon~\cite{Carr:1974nx,Carr:1975qj}. The masses of the PBHs follow the critical scaling law~\cite{Choptuik:1992jv,Niemeyer:1997mt,Niemeyer:1999ak}
\be \label{eq:Mcrit}
    M(\delta) = \kappa M_k (\delta - \delta_c)^\gamma \,,
\ee
where $M_k = M_p^2/(2H(t_k)) = M_0 H_0/H(t_k)$ denotes the horizon mass when the scale $k$ re-enters horizon. The horizon mass during the inflation can be approximated as 
\be
    M_0 \approx 0.05 M_\odot \left[ \frac{100}{g_*} \right]^{\!\frac12} \left[\frac{T_{\rm reh}}{\rm GeV}\right]^{-2} \,.
\ee
The parameters $\gamma$, $\kappa$ and $\delta_c$ depend on the sphericity and the profile of the overdensity as well as the equation of state of the Universe~\cite{Musco:2018rwt,Young:2019yug,Musco:2020jjb,Yoo:2020lmg,Franciolini:2022tfm,Musco:2023dak}. We use the fixed values $\gamma = 0.38$, $\kappa = 4.2$ and $\delta_c = 0.55$, corresponding to the perfect radiation-fluid case~\cite{Franciolini:2022tfm}.

The fraction of the total energy density that collapses to PBHs of mass $M$ is given by~\cite{Carr:1975qj,Gow:2020bzo}
\bea
    \beta_k(M) &= \int_{\delta_c} \td \delta \frac{M}{M_k} P_k(\delta) \,\delta_D\!\left(\ln \frac{M}{M(\delta)}\right) \\
    &= \frac{\kappa}{\gamma} \left(\frac{M}{\kappa M_k}\right)^{\!1+\frac{1}{\gamma}} P_k(\delta(M)) \,,
\eea
where $\delta_D$ denotes the Dirac delta function and $\delta(M) = \delta_c + [M/(\kappa M_k)]^{1/\gamma}$. From $\beta_k(M)$ we obtain the present PBH mass function as
\bea
    &\psi(M) = \int \!\td \ln k \, \beta_k(M) \frac{\rho_r(T_k)}{\rho_c} \frac{s(T_0)}{s(T_k)} \\
    &\approx \frac{2.0\times 10^8}{h^2} \frac{g_*}{g_{*s}} \frac{T_{\rm reh}}{\rm GeV} \!\int \!\td \ln k\, \beta_k(M)  \bigg[\frac{H(t_k)}{H_0}\bigg]^{\frac12} ,
\eea
where $\rho_c$ denotes the critical energy density of the Universe, $s(T)$ the entropy density, $T_k$ the temperature at the horizon re-entry of the scale $k$, $T_0$ the present CMB temperature and in the last step we approximated that $g_*$ and $g_{*s}$ don't significantly change when the relevant scales re-enter horizon. The total PBH abundance is $\Omega_{\rm PBH} = \int \td \ln M \,\psi(M)$.

\begin{figure}
\centering
\includegraphics[width=\columnwidth]{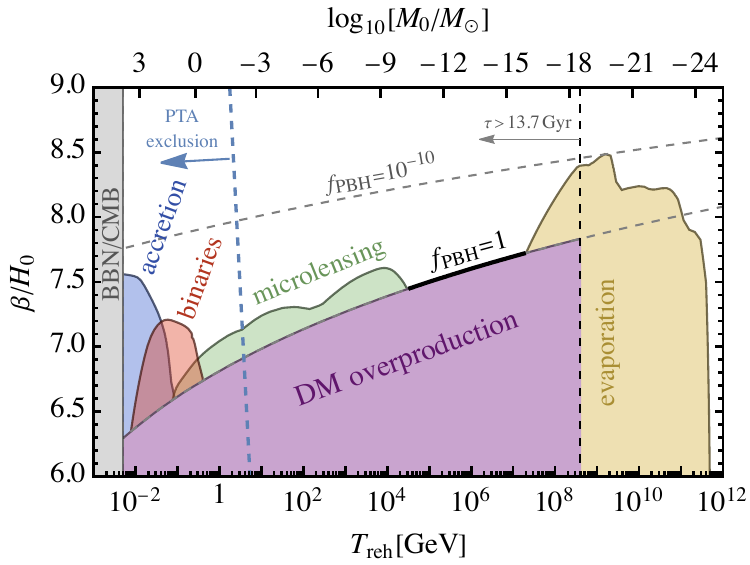}
\caption{The PBH constraints projected to the plane of the PT parameters. The gray region is excluded by the BBN/CMB constraint on the reheating temperature and the region left from the dashed blue curve by the PTA observations.}
\label{fig:PBHplot}
\end{figure}

\begin{figure}
\centering
\includegraphics[width=0.97\columnwidth]{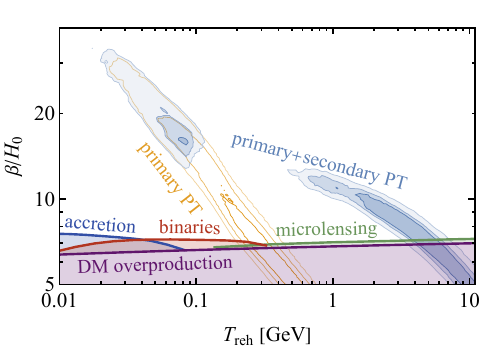}
\caption{The blue contours show the fit of the spectrum including the primary and secondary contributions to the recent NANOGrav data. The orange contours show the same fit using only the primary spectrum. The coloured regions at the bottom are excluded by constraints on the PBHs produced in the transition.}
\label{fig:PTAfit}
\end{figure}

We compute the PBH mass function and abundance using the fits~\eqref{eq:Pkfit} of $P_k(\delta)$. By performing the computation for several $\beta/H_0$ values and doing numerical fits, we find that the fraction of DM in PBHs, $f_{\rm PBH}\equiv \Omega_{\rm PBH}/\Omega_{\rm DM}$, is roughly of the form 
\be \label{eq:fPBH}
    f_{\rm PBH} = b_1 \exp\!\left[-b_2 e^{b_3 \beta/H_0}\right] \frac{g_*}{g_{*s}} \frac{T_{\rm reh}}{\rm GeV}
\ee
with $b_1 \approx 5.5\times 10^6$, $b_2 \approx 0.064$ and $b_3 \approx 0.806$~\cite{SM}, and the mass function is of form typical for the critical scaling law~\cite{Vaskonen:2020lbd}, 
\be \label{eq:psifit}
    \psi(M) \propto (M/M_0)^{1+1/\gamma} \exp[-c_1 (M/M_0)^{c_2}] \,.
\ee
The values of $c_1$ and $c_2$ depend mildly on $\beta/H_0$. For $\beta/H_0 \approx 7$ we get $c_1\approx 1.2$ and $c_2 \approx 2.7$. The abundance of PBHs is constrained by various observations: the evaporation effects on BBN and the extragalactic gamma-ray background~\cite{Carr:2009jm, Acharya:2020jbv}, lensing of stellar light~\cite{Griest:2013aaa,Niikura:2017zjd,Smyth:2019whb,EROS-2:2006ryy,Niikura:2019kqi,Macho:2000nvd}, supernovae~\cite{Zumalacarregui:2017qqd} and GWs~\cite{Urrutia:2021qak,Urrutia:2023mtk}, the binary merger rate~\cite{Hutsi:2020sol}, survival of stars in dwarf galaxies~\cite{Brandt:2016aco,Koushiappas:2017chw}, survival of wide binaries~\cite{Monroy-Rodriguez:2014ula}, Lyman-$\alpha$ forest~\cite{Afshordi:2003zb,Murgia:2019duy} 
and limits on radiation emitted in accretion process~\cite{Ricotti:2007au, Horowitz:2016lib,Ali-Haimoud:2016mbv, Poulin:2017bwe, Hektor:2018qqw, Hutsi:2019hlw, Serpico:2020ehh}. In Fig.~\ref{fig:PBHplot} we show the projection of the PBH constraints in the plane of the PT parameters calculated using the fits~\eqref{eq:fPBH} and~\eqref{eq:psifit} and the method of~\cite{Carr:2017jsz}. PBHs can constitute all DM in the asteroid mass window corresponding to $3\times 10^4 \lesssim T_{\rm reh}/{\rm GeV} \lesssim 2\times 10^7$ and $7.4 \lesssim \beta/H_0 \lesssim 7.7$. The gray region on the left is excluded by the BBN/CMB constraint $T_{\rm reh} > 5\,$MeV~\cite{deSalas:2015glj,Hasegawa:2019jsa,Allahverdi:2020bys} and the region left from the blue dashed curve by the NANOGrav observations (see Fig.~\ref{fig:PTAfit}).

Compared to~\cite{Liu:2021svg, Kawana:2022olo, Gouttenoire:2023naa, Baldes:2023rqv}, our PBH computation includes two major improvements: 1) We account accurately for the distributions of the bubble nucleation times in the computation of the evolution of the false vacuum fraction while in~\cite{Liu:2021svg, Kawana:2022olo, Gouttenoire:2023naa, Baldes:2023rqv} the fluctuations were only in the time when the nucleation begins. 2) We compute the perturbations as a function of $k\leq k_{\rm max}$ while in~\cite{Liu:2021svg, Kawana:2022olo, Gouttenoire:2023naa, Baldes:2023rqv} only the scale of maximal $\delta$ was considered. Combined with incorporation of the critical scaling law~\eqref{eq:Mcrit}, these improvements result in a more realistic estimate of the PBH mass function and abundance. 

The main limitation of our computation is that the subhorizon inhomogeneities are still large and the Universe is not yet fully radiation-dominated when the smallest scales $k\simeq k_{\rm max}$ re-enter horizon. Further work is required to ascertain their collapse. However, we have estimated that the assumption of radiation dominance holds already at $k = 0.9k_{\rm max}$ for $\beta/H_0 > 7$ and cutting off the scales for which the subhorizon inhomogeneities can be relevant results only in a small shift in $\beta/H_0$~\cite{SM}.

We note that our result~\eqref{eq:fPBH} is based on fitting the distributions of $\delta$ obtained by generating $10^6$ realizations of the patches. However, we have checked that reducing the number of realizations to $10^5$ changes the result for the PBH abundance only by $\mathcal{O}(10\%)$. Moreover, this uncertainty does not affect the secondary GWs discussed in the following as they are not sensitive to the tails of the distribution.

\vspace{5pt}\noindent\textbf{Gravitational waves --} For strongly supercooled PTs there are two distinct sources of GWs: the collisions of the bubble walls/relativistic fluid shells~\cite{Kosowsky:1992vn,Lewicki:2022pdb} and the large curvature perturbations that induce production of GWs at second order~\cite{Tomita:1975kj,Matarrese:1993zf,Mollerach:2003nq,Ananda:2006af,Baumann:2007zm,Acquaviva:2002ud,Domenech:2021ztg}. 

Various primary spectra (PGW) from very strong transitions have been predicted in the literature depending on the model and the approximations used~\cite{Kosowsky:1992vn,Huber:2008hg,Weir:2016tov,Jinno:2016vai,Jinno:2017fby,Konstandin:2017sat,Cutting:2018tjt,Lewicki:2020jiv,Lewicki:2020azd,Cutting:2020nla,Lewicki:2022pdb}. We incorporate the semi-analytical modelling describing bubble wall collisions and relativistic fluid shells developed in~\cite{Lewicki:2020azd,Lewicki:2022pdb} which was used e.g. in~\cite{Franciolini:2023wjm,Blanco-Pillado:2024aca}. The spectrum is a broken power-law
\bea \label{eq:Omega_PT}
    \Omega_{\rm PGW} = \left[\frac{\beta}{H}\right]^{-2} \!\!\frac{A (a+b)^c S_H(k,k_{\rm max})}{\left[b \left({k}/{k_p}\right)^{\!-\frac{a}{c}} + a \left({k}/{k_p}\right)^{\!\frac{b}{c}}\right]^{\!c}} \,,
\eea
where $k_p \approx 0.7 k_{\rm max} \beta/H_0$, $A=5.1\times 10^{-2}$, $a=b=2.4$, $c=4$~\cite{Lewicki:2022pdb} and 
\be \label{eq:SH}
    S_H(k,k_{\rm max}) = \left[1 + \frac{\Omega_{\rm CT}(k_{\rm max})}{\Omega_{\rm CT}(k)}\left(\frac{k}{k_{\rm max}}\right)^{\!a}\right]^{-1} \,.
\ee
The function $\Omega_{\rm CT}(k)$ accounts for the causality-limited tail of the spectrum at $k\lesssim k_{\rm max}$. In pure radiation dominance $\Omega_{\rm CT}(k) \propto k^3$~\cite{Caprini:2009fx} but becomes slightly less steep e.g. around the QCD transition~\cite{Franciolini:2023wjm}. The results we use for the primary GW spectrum are obtained from simulations neglecting the expansion of the Universe. Including it may lead to an additional suppression of the primary spectrum~\cite{Zhong:2021hgo}.

The spectrum of the secondary GWs (SGWs) reads~\cite{Kohri:2018awv,Espinosa:2018eve,Inomata:2019yww}
\bea
    \Omega_{\rm SGW} \approx \frac{1}{3} \int_1^\infty \!\!\td t & \int_{0}^{1} \!\td s \,\mathcal{I}_{t,s}^2  \left [ \frac{(t^2-1)(1-s^2)}{t^2-s^2} \right ]^2 \\
    &\times \mathcal{P}_\zeta\left(k\frac{t-s}{2}\right) \mathcal{P}_\zeta\left(k\frac{t+s}{2}\right) ,
\eea
where $\mathcal{P}_\zeta$ denotes the curvature power spectrum and $\mathcal{I}_{t,s}$ the transfer function,
\bea \label{eq:tranfer}
    {\cal I}_{t,s}^2 & = \!\frac{288(s^2+t^2-6)^2}{(t^2-s^2)^6} \!\Bigg[ \!\frac{\pi^2}{4} (s^2+t^2-6)^2 \theta (t^2-3) \\
    & + \left(t^2 - s^2 - \frac{1}{2} (s^2+t^2-6) \ln \left| \frac{t^2-3}{3-s^2} \right| \right)^2 \Bigg] .
\eea
We neglect the corrections arising from the non-Gaussian shape of the perturbations as the abundance of the induced GWs is mainly determined by the characteristic amplitude of perturbations. This slightly underestimates the SGW abundance~\cite{Unal:2018yaa,Cai:2018dig,Yuan:2020iwf,Adshead:2021hnm,Abe:2022xur,Ellis:2023oxs}. Following the same reasoning, we use the linear relation between the density contrast and the curvature perturbation at the time of horizon crossing, $\zeta = 9 \delta/4$, to compute $\mathcal{P}_\zeta$ from the spectrum of $\delta$. The assumption of radiation domination during horizon re-entry needed for this relation, as well as for the transfer function~\eqref{eq:tranfer}, holds already at $k>0.9k_{\rm max}$. The relevant scales are, however, not at any point very deep superhorizon. We leave studies of this issue for future work.

\begin{figure}
\centering
\includegraphics[width=\columnwidth]{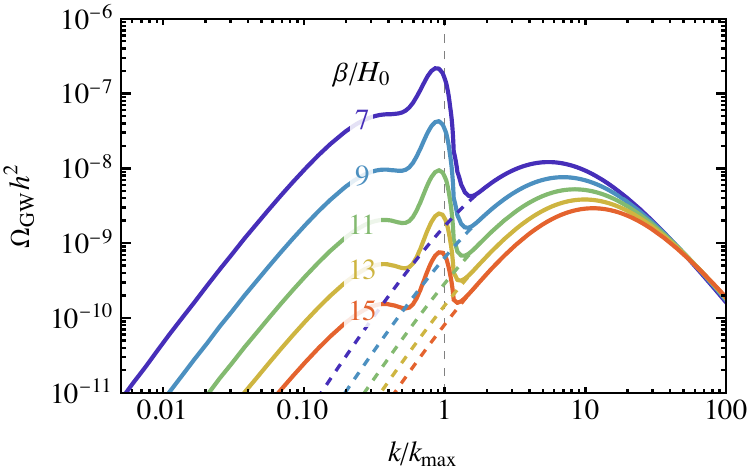}
\caption{The solid curves show the sum of the primary and secondary GW spectra from a PT as a function of the wavenumber $k=2\pi f$. The dashed curves show the primary spectrum alone.}
\label{fig:OmegaGW}
\end{figure}

In Fig.~\ref{fig:OmegaGW} we show the total present abundance of the GWs produced in the PT,
\be
    \Omega_{\rm GW} h^2 \approx 1.6\times 10^{-5} \!\left[ \frac{g_*}{100} \right] \!\left[ \frac{g_{*s}}{100} \right]^{\!-\frac43} \left[ \Omega_{\rm PGW} + \Omega_{\rm SGW} \right] ,
\ee 
for different $\beta/H_0$ values. The spectrum has two peaks corresponding to the two sources. The SGW spectrum exhibits a low-frequency tail that scales roughly as $\propto k^{2.5}$, a narrow peak at $k\simeq k_{\rm max}$ and a sharp fall-off at $k > k_{\rm max}$. The peak amplitude of the secondary spectrum can be approximated by
\be
    \Omega_{\rm SGW}|_{\rm peak} = a_1 e^{- a_2 \beta/H_0} ,
\ee
with $a_1 \approx 2.0$ and $a_2 \approx 0.75$~\cite{SM}, while its shape does not change very significantly with $\beta/H_0$. The PGW spectrum dominates at $k > k_{\rm max}$ and has its maximum at $k_p > k_{\rm max}$. As the peak amplitude of the PGW spectrum scales only as a power-law, $\Omega_{\rm PGW} \propto (\beta/H_0)^{-2}$, it becomes dominant at large $\beta/H_0$. The peak amplitudes of the primary and secondary GW spectra are the same for $\beta/H_0 \approx 12$.

Recently, the pulsar timing array collaborations have reported strong evidence for a stochastic GW background at nHz frequencies~\cite{NANOGrav:2023gor, NANOGrav:2023hde, EPTA:2023fyk, EPTA:2023sfo, EPTA:2023xxk, Reardon:2023gzh, Zic:2023gta, Reardon:2023zen, Xu:2023wog, NANOGrav:2023hvm,  InternationalPulsarTimingArray:2023mzf}. The primary spectrum from a strong PT was already featured in a multi-model analysis in~\cite{Ellis:2023oxs}. In Fig.~\ref{fig:PTAfit} we show an improved fit including both the spectra. The inclusion of the secondary spectrum is crucial and the improved reconstructed parameter space is not compatible with the previous results. In particular, the best-fit point moves to significantly higher temperatures, $T_{\rm reh}=2.6\,$GeV with $\beta/H_0=8.6$. However, the quality of the fit changes very marginally by $\Delta \chi^2\approx 0.08$. The negative non-Gaussianity of the perturbations is important here, as otherwise the second-order GWs are disfavoured by the PTA fit due to the PBH overproduction~\cite{Franciolini:2023pbf}.

\begin{figure}
\centering
\includegraphics[width=0.98\columnwidth]{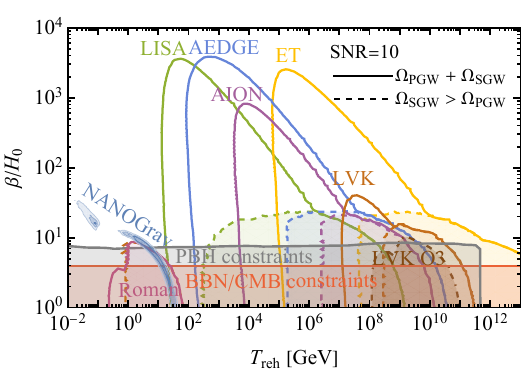}
\caption{The solid curves show the parts of the parameter space within reach of indicated experiments with ${\rm SNR}=10$. The dashed curves indicate the parts where ${\rm SNR}=10$ is reached by integrating only over the frequency range where the secondary spectrum dominates. The gray region is excluded by the PBH constraints shown in Fig.~\ref{fig:PBHplot} and the red region by CMB and BBN constraints on the GW abundance. The blue area corresponds to the fit to NANOGrav data from Fig.~\ref{fig:PTAfit}.}
\label{fig:SNRs}
\end{figure}

In Fig.~\ref{fig:SNRs} we show the regions of the PT parameters that planned GW experiments will be able to probe with signal-to-noise ratio ${\rm SNR} \geq 10$. Solid lines show where the sum of the primary and secondary PT spectra will be probed while the dashed ones indicate only regions where the secondary spectrum dominates and will be detectable. We include the reach of LISA~\cite{Bartolo:2016ami, Caprini:2019pxz, LISACosmologyWorkingGroup:2022jok}, AEDGE~\cite{AEDGE:2019nxb,Badurina:2021rgt}, AION~\cite{Badurina:2019hst,Badurina:2021rgt}, ET~\cite{Punturo:2010zz, Hild:2010id}, the Nancy Roman telescope~\cite{Wang:2022sxn}, and LVK through the design sensitivity of 
LIGO~\cite{LIGOScientific:2014pky, LIGOScientific:2016fpe, LIGOScientific:2019vic}. The region below the red line is excluded by the BBN/CMB constraint on the total abundance of GWs, $\Omega_{\rm GW} h^2 \lesssim 1.8\times 10^{-6}$~\cite{Kohri:2018awv, Pagano:2015hma} and the dark brown dot-dashed line is the constraint from the current LVK (O3) data~\cite{KAGRA:2021kbb}. The gray region shows where the produced population of PBHs is constrained. While the primary spectrum dominates for fast transitions, for slow ones with $\beta/H_0 \lesssim 70$, the secondary spectrum will also be detectable. 
The region where the secondary GWs would be observable also extends to relatively high reheating temperatures where only the low-frequency tail, falling off slower for the secondary contribution than for the primary one, is within reach.
Remarkably, the current LVK constraint reaches above the PBH constraints. This is mainly because of the secondary GW component and the suppression of the PBH formation arising from the negative non-Gaussianity.

\vspace{5pt}\noindent\textbf{Conclusions --} In this letter we have studied the large inhomogeneities generated in slow, strongly supercooled first-order PTs. We have derived a new contribution to the GW spectrum, the secondary GWs, resulting from the PT. The process behind this contribution is identical to the well-known scalar-induced GWs associated with the primordial inflation. We have shown that the secondary GWs dominate the spectrum and have a very significant impact on its shape, provided the transition is slow enough.

We have computed also the distribution of the perturbations finding it to have a negative non-Gaussianity. This has allowed us to better estimate the formation of PBHs in PTs. We have shown that models with a PT occurring at $T_{\rm reh} \approx 10^6\,$GeV with $\beta/H_0 \approx 7.5$ would provide the entirety of the observed dark matter in the form of asteroid mass PBHs and produce a strong GW spectrum within the reach of LISA, ET, AION and AEDGE. The smoking gun signature of our scenario is the peculiar shape of the GW spectrum, including two peaks corresponding to the horizon scale at the end of the transition and the characteristic size of the bubbles.

\begin{acknowledgments}
\vspace{5pt}\noindent\emph{Acknowledgments --} We would like to thank Yann Gouttenoire for the helpful discussions and the organizers of the workshop {\it Early Universe cosmology with Gravitational Waves and Primordial Black Holes} at the University of Warsaw where this work was initiated. The work of M.L. and P.T. was supported by the Polish National Agency for Academic Exchange within the Polish Returns Programme under agreement PPN/PPO/2020/1/00013/U/00001 and the Polish National Science Center grant 2018/31/D/ST2/02048. The work of V.V. was supported by the European Union's Horizon Europe research and innovation program under the Marie Sk\l{}odowska-Curie grant agreement No. 101065736, and by the Estonian Research Council grants PRG803, RVTT3 and RVTT7 and the Center of Excellence program TK202.
\end{acknowledgments}

\bibliography{refs}

\newpage
\clearpage
\onecolumngrid

\begin{center}
\textbf{\large \papertitle} \\ 
\vspace{0.06in}
{Marek Lewicki, Piotr Toczek and Ville Vaskonen} \\ 
\vspace{0.1in}
{SUPPLEMENTAL MATERIAL}
\vspace{0.1in}
\end{center}

\setcounter{equation}{0}
\setcounter{figure}{0}
\setcounter{section}{0}
\setcounter{table}{0}
\setcounter{page}{1}
\makeatletter
\renewcommand{\theequation}{S\arabic{equation}}
\renewcommand{\thefigure}{S\arabic{figure}}
\renewcommand{\thetable}{S\arabic{table}}

\twocolumngrid

\section{Computation of \\ the false vacuum fraction} 

The average false vacuum fraction is given by the probability that the given point still is in the false vacuum at time $t$:
\be \label{eq:Fbar}
	\bar{F}(t) = \exp\left[-\frac{4\pi}{3} \int_{-\infty}^t \! \td t_n\, \Gamma(t_n) a(t_n)^3 R(t,t_n)^3\right] \,.
\ee
Here $R(t_n,t)$ is the comoving radius of a bubble nucleated at time $t_n$. The bubble radius increases linearly with the conformal time. In a finite volume, the evolution of the false vacuum fraction can deviate significantly from $\bar{F}(t)$. In the following, we derive a semi-analytical way of computing the false vacuum fraction in a finite comoving volume.

\begin{figure}
\centering
\includegraphics[width=0.94\columnwidth]{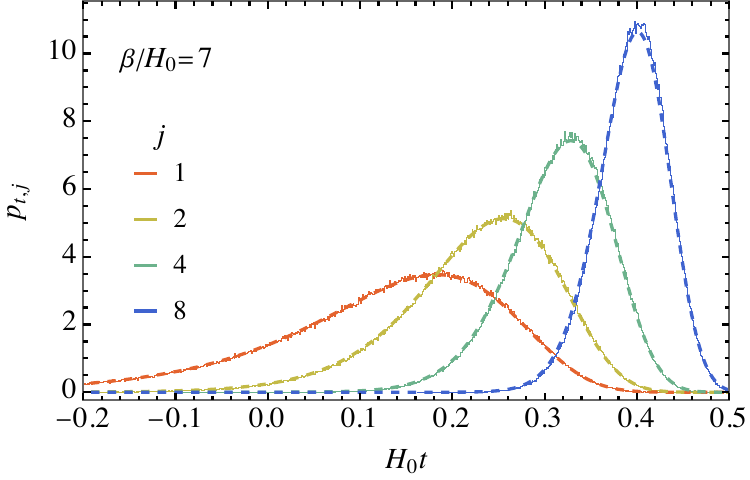}
\caption{The probability distributions of the nucleation times of the 1st, 2nd, 4th and 8th bubbles inside the volume $V(0.9 k_{\rm max})$. The thin histograms result from $10^6$ numerical simulations and the dashed curves show the function~\eqref{eq:pt}.}
\label{fig:ptj}
\end{figure}

Consider a spherical comoving volume $V(k) = 4\pi k^{-3}/3$. We divide the false vacuum fraction $F_k(t)$ in the volume $V(k)$ into two pieces: the contribution from the first $j_c$ bubbles and the contribution from the rest of the bubbles, 
\be
    F_{k}(t) = F_{k}^{(j\leq j_c)}(t) F_{k}^{(j > j_c)}(t) \,.
\ee
As the bubble nucleation is a Poisson process, the probability distribution of the time when the $j$th bubble reaches or nucleates within that volume is given by
\be \label{eq:pt}
	p_{t,j}(t;k) = \frac{\td \bar{N}_k(t)}{\td t} \frac{\bar{N}_k(t)^{j-1}}{\Gamma(j)} e^{-\bar{N}_k(t)} \,,
\ee
where
\be
	\bar{N}_k(t) = \frac{4\pi}{3} \int_{-\infty}^t \td t_n \, \Gamma(t_n) a(t_n)^3 \left[ k^{-1} + R(t;t_n) \right]^3
\ee
is the expected number of bubbles that nucleate inside or reach the volume $a^3 V(k)$ by the time $t$. We show the distributions $p_{t,j}$ in Fig~\ref{fig:ptj} for $j=1,2,4,8$, $k=0.9 k_{\rm max}$ and $\beta/H_0 = 7$.

The bubble that at time $t$ reaches or nucleates inside the volume $V(k)$ must have nucleated within the distance $d < k^{-1} + R(t;t_n)$ where $t_n$ is its nucleation time. The probability distribution of $d$ increases as $d^2$ up to that maximal radius:
\be \label{eq:pd}
    p_d(d;t,k) = \frac{4\pi d^2}{\bar{N}_k(t)} \int_{-\infty}^t \!\!\!\td t_n \Gamma(t_n) a(t_n)^3 \theta(k^{-1} + R(t;t_n) - d)
\ee
The volume of the intersection of spheres of radius $R$ and $r$ whose centres are separated by distance $d$ is
\be
    V_{\rm int}(d,R,r) \!=\! 
    \begin{cases}
	\frac{4\pi}{3} d^3 \xi(\frac{R}{d},\frac{r}{d}) & |R-r| \leq d \leq R+r \\
	\frac{4\pi}{3} \min[r,R]^3 & d < |R-r| \\
	0 & d > R+r
	\end{cases} ,
\ee
where $\xi(x,y) = (1-x-y)^2(1+2(x+y)-3(x-y)^2)/16$. So, the $j$th bubble covers the fraction
\be \label{eq:f}
	f(t;t_j,d_j,k) \equiv \frac{V_{\rm int}(d_j,\max[0,d_j-k^{-1}]+R(t;t_j),k^{-1})}{V(k)}
\ee
of the volume $V(k)$. 

The contribution from the first $j_c$ bubbles can be computed either by numerical simulations where these bubbles are nucleated inside some volume according to the nucleation rate $\Gamma(t)$ or semi-analytically by generating times $t_j$ when the $j$th bubble nucleates or reaches the volume $V(k)$ from the rate $\td \bar{N}_k/\td t$ and picking random numbers from the distribution $p_d$ for the distance at which the $j$th bubble nucleates.\footnote{Generating the times $t_j$ from $\td \bar{N}_k/\td t$ is numerically very fast and the resulting distribution of $t_j$ is $p_{t,j}$. Notice that, picking the times $t_j$ from the distributions $p_{t,j}$ would allow $t_j'< t_j$ for $j'>j$ and after ordering the times $t_j$ the resulting distributions would not match $p_{t,j}$.} The function $F_k^{(j<j_c)}$ can be simply integrated from the numerical simulations whereas in the semi-analytical approach, it can be approximated as 
\be \label{eq:Fkstrong}
    F_{k}^{(j\leq j_c)}(t) \approx \prod_{j=1}^{j_c} \left[1-f(t;t_j,d_j,k)\right] \,.
\ee
For the main results of the paper, we use the semi-analytical approach which is computationally less expensive allowing large scans of the parameters including the wavenumber $k$ and the inverse time scale of the transition $\beta$. We have, however, performed cross-checks of the results with numerical simulations and found a very good agreement between the approaches. In particular, as shown in Fig.~\ref{fig:ptj}, the probability distribution of the bubble nucleation times~\eqref{eq:pt}, matches with the results of numerical simulations.

\begin{figure}
\centering
\includegraphics[width=0.94\columnwidth]{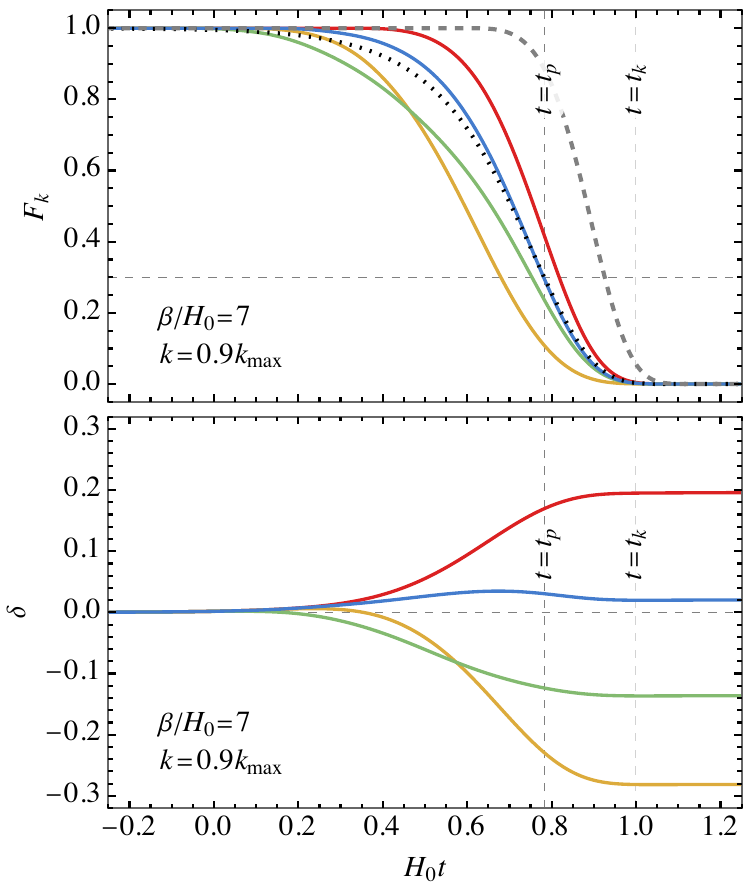}
\caption{\emph{Upper panel:} The colourful curves represent the false vacuum fraction in four patches of radius $1/(0.9 k_{\rm max})$ and the dotted black curve shows the time evolution of the expected false vacuum fraction $\bar F(t)$. The dashed gray curve shows $F_{k}^{(j>j_c)}(t)$ for $j_c = 50$. \emph{Lower panel:} The evolution of the density contrast in the same four patches as shown in the upper panel. The vertical dashed lines indicate the percolation time $t_p$ and the time $t_k$ when the scales associated with these patches re-enter the horizon.}
\label{fig:Fkt}
\end{figure}

The contribution from the rest of the bubbles can be approximated as
\bea \label{eq:Fkweak}
	F_{k}^{(j>j_c)}(t) \approx \exp\!\bigg[-&\!\!\!\!\sum_{j=j_c+1}^\infty \int  \td t_j \td d_j \,p_{t,j}(t_j;k) \\
    &\times p_d(d_j;t_j,k) f(t;t_j,d_j,k) \bigg] .
\eea
Picking a large enough $j_c$, we approximate the distribution~\eqref{eq:pt} for $j>j_c$ by the delta function
\be
    p_{t,j}(t;k) = \frac{\td \bar{N}_k(t)}{\td t} \delta(\bar{N}_k(t)-j)
\ee 
and the sum over $j$ by an integral. Then, sum in the exponent in~\eqref{eq:Fkweak} reduces to a simple two-dimensional integral:
\bea \label{eq:FJ}
    &\sum_{j=j_c + 1}^\infty \int \td t_j \td d_j \,p_{t,j}(t_j;k)  p_d(d_j;t_j,k) f(t;t_j,d_j,k) \\
    &\approx \int\! \td t' \td d' \,\theta\!\left( \bar{N}_k(t') - j_c \right)\! \frac{\td \bar{N}_k(t')}{\td t'} p_d(d';t',k) f(t;t',d',k) .
\eea
We plot the function $F_{k}^{(j>j_c)}(t)$ for $j_c = 50$ by the gray dashed curve in the upper panel of Fig.~\ref{fig:Fkt}.

The function $F_{k}^{(j>j_c)}(t)$ approaches the expected evolution $\bar{F}(t)$ at small $k$. This can be shown analytically: Taking a large enough volume ($k^{-1}$ much larger than the average bubble radius), the bubbles nucleating outside the volume can be neglected. In this limit, we can use the approximations
\be
	\bar{N}_k(t) \approx \frac{4\pi}{3} k^{-3} \int_{-\infty}^t \td t_n \, \Gamma(t_n) a(t_n)^3 \,,
\ee
and
\be
    f(t;t_j,d_j,k) \approx (kR(t;t_j))^3 \,,
\ee
which give
\bea
	F_{k}^{(j>j_c)}(t) \approx \exp\!\bigg[ - \frac{4\pi}{3} &\int \!\td t' \theta\!\left( \bar{N}_k(t') - j_c \right) \\ &\times \Gamma(t') a(t')^3 R(t;t')^3  \bigg] .
\eea
The $j_c\to 0$ limit or, equivalently, the $k\to 0$ limit, as in that case the first few bubbles can be neglected, matches the average evolution~\eqref{eq:Fbar}, $\lim_{j_c\to 0} F_{k}(t) = \lim_{k\to 0} F_{k}(t)  = \bar{F}(t)$.

\section{Computation of the density contrast} 

We show the evolution of the false vacuum fraction in four patches corresponding to the wavenumber $k = 0.9 k_{\rm max}$ in the upper panel of Fig.~\ref{fig:Fkt}. These are computed with $j_c = 50$. From $F_k(t)$ we can compute the density contrast 
\be
    \delta(t) = \frac{\rho_k(t)}{\bar\rho(t)} - 1\,,
\ee
by solving the total energy density from the continuity equation describing the production of radiation from the decaying false vacuum in an expanding universe. We show the density contrast as a function of time in the lower panel of Fig.~\ref{fig:Fkt} for the same patches as in the upper panel. Note that from this plot we see that the assumption of radiation dominance used for the PBH computations in the main text holds at the horizon reentry of $k = 0.9 k_{\rm max}$ for $\beta/H_0 = 7$.

\begin{figure}
\centering
\includegraphics[width=0.94\columnwidth]{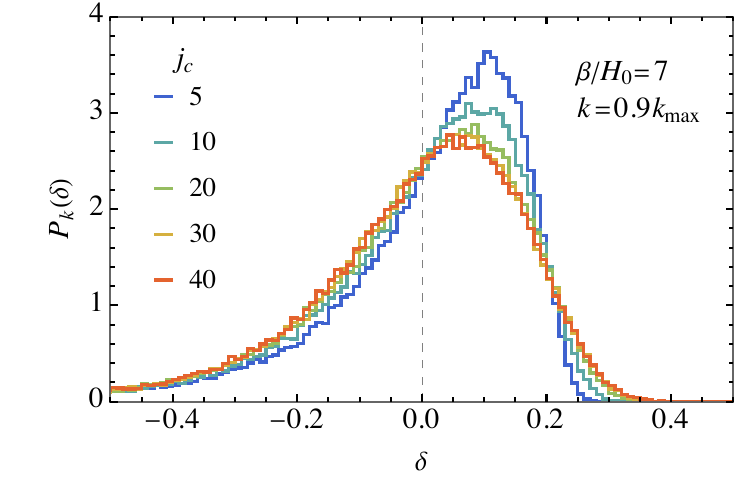}
\caption{The probability distribution of the density contrast for different choices of the cut-off $j_c$.}
\label{fig:Pkdeltajc}
\end{figure}

In the main text, we use the distribution of $\delta$ at the horizon re-entry, $t=t_k$, to estimate the PBH and GW abundances generated in the PT. Here, we investigate how the distribution of $\delta$ depends on the choice of $j_c$. We generate $10^5$ realizations of $F_k^{j\leq j_c}(t)$ for $k=0.9k_{\rm max}$, $\beta/H_0 = 7$ and different choices of the cut-off: $j_c = 5,\,10,\,20,\,30,\,40$\,. For each resulting $F_k(t)$ we compute $\delta(t)$ and evaluate them at the percolation temperature $t_p$ to obtain the distribution of $\delta(t_p)$ that we denote by $P_k(\delta)$. As shown in Fig.~\ref{fig:Pkdeltajc}, too small choices of $j_c$ underestimate the width of the distribution but for $j_c \geq 30$ the distribution remains almost unchanged. For our main results, we use $j_c = 50$. With our implementation using \texttt{C++}, the computation of $10^6$ realizations (for fixed $\beta/H_0$ and $k$) takes about 30\,minutes on an Apple M1 Pro 8-core processor. The code is available at \href{https://github.com/vianvask/deltaPT}{https://github.com/vianvask/deltaPT}.

\section{Fits of the PBH abundance \\ and the SGW amplitude} 

\begin{figure}
\centering
\includegraphics[width=0.94\columnwidth]{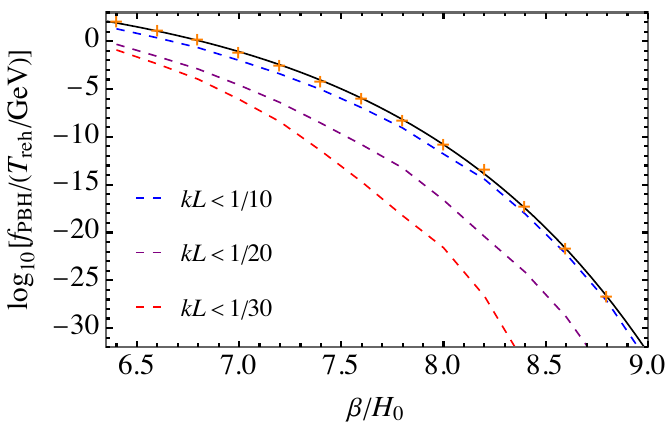}
\includegraphics[width=0.94\columnwidth]{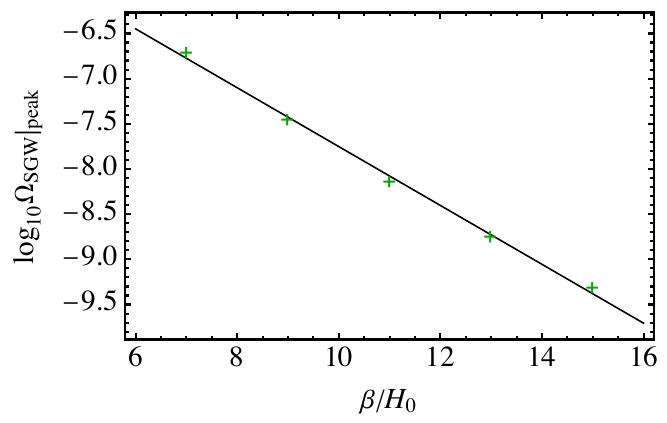}
\caption{The solid curves show the fits of the PBH abundance and the SGW peak amplitude obtained, respectively, using the fitting functions~\eqref{eq:PBHfit} and~\eqref{eq:GWfit}. In the upper panel, the numerical results marked in orange are obtained by integrating up to $k=k_{\rm max}$. For comparison, the dashed lines show the result if the smallest length scales, for which the scale of the inhomogeneities exceed a given bound, are not included.}
\label{fig:fits}
\end{figure}

For convenience, we provide fits of the numerical results of the PBH abundance and the SGW peak amplitude in the main text. We use a double exponential fitting function for the PBH abundance,
\be \label{eq:PBHfit}
    f_{\rm PBH} = b_1 \exp\!\left[-b_2 e^{b_3 \beta/H_0}\right] \frac{g_*}{g_{*s}} \frac{T_{\rm reh}}{\rm GeV} \,,
\ee
and an exponential fitting function for the SGW peak amplitude,
\be \label{eq:GWfit}
    \Omega_{\rm SGW}|_{\rm peak} = a_1 e^{-a_2 \beta/H_0} \,.
\ee
We find that the variance of the density fluctuations scales roughly exponentially with $\beta/H_0$ and the fitting functions reflect the expectation that the PBH abundance scales exponentially and the GW amplitude linearly with the variance. As shown in Fig.~\ref{fig:fits}, these fitting functions provide good fits of the numerical results in the relevant ranges of $\beta/H_0$. The best fit parameter values and the standard errors are given by $b_1 = (5.5\pm 4.0)\times 10^6$, $b_2 = 0.064\pm 0.010$ and $b_3 = 0.806 \pm 0.017$ for the PBH abundance fit and by $a_1 = (3.2 \pm 0.9)\times 10^{-5}$, $a_2 = 0.75 \pm 0.03$ for the SGW amplitude fit.

In the upper panel of Fig.~\ref{fig:fits} we illustrate also the uncertainty of the PBH abundance arising from the potential suppression of the PBH formation caused by subhorizon inhomogeneities. These arise because, for some time, most of the energy after the bubble collisions remains in thin shells that keep propagating with the speed of light. We estimate the scale of the subhorizon inhomogeneities as $L \equiv 3V_k/S_{\rm tot}$ where
\bea
    S_{\rm tot} =& \,4\pi \int_{-\infty}^{t_k} \td t_n \,\bar{F}(t_n) \Gamma(t_n) a(t_n)^3  \\
    &\times \int_0^{k^{-1} + R(t_k;t_n)} \! \td d \, d^2 S_{\rm int}(d, R(t_k;t_n), k^{-1}) \,,
\eea
is the area of the shells at the moment when the scale $k$ reenters horizon. The function $S_{\rm int}(d,R,r)$ gives the surface area of the sphere of radius $R$ that is enclosed within a sphere of radius $r$, with the distance $d$ between the spheres:
\be
    S_{\rm int}(d,R,r) \!=\!
    \begin{cases}
	\frac{\pi R \left[r^2 - (R-d)^2\right]}{d} & |R-r| \leq d \leq R+r \\
	4\pi R^2 & d < r-R \\
	0 & d > R+r \vee d < R-r
	\end{cases} .
\ee 
If the scale of the inhomogeneities is very small compared to the radius of the patch, $kL \ll 1$, we expect that they don't significantly affect the collapse. Putting an upper bound on $kL$ induces an upper bound on $k$ below which the PBH formation proceeds roughly as in the case of a smooth horizon scale overdensity. Determining how strongly the subhorizon inhomogeneities affect the PBH formation and how strong upper bound that would put on $kL$ would require lattice simulations that are beyond the scope of our current work. However, from Fig.~\ref{fig:fits} it is evident that, since the PBH abundance increases very rapidly with decreasing values of $\beta/H_0$, neglecting the largest wavenumbers that exit horizon during the thermal inflation results only in a small shift in the relevant values of $\beta/H_0$.

\end{document}